%% file: main.tex
\documentclass[copyright]{eptcs}
\providecommand{\event}{RTRTS 2010} % Name of the event you are submitting to
\usepackage{breakurl}             % Not needed if you use pdflatex only.

\usepackage{latexsym}
\usepackage{alltt}
\usepackage{epsfig}
\usepackage{url}
\usepackage{amsfonts}
\usepackage{graphicx}
\usepackage{amssymb}
\usepackage{float}

\usepackage{shortvrb}  % Ollie stuff
\usepackage{dsfont}    % Ollie stuff
 \MakeShortVerb{\@}     % Ollie stuff!

\title{Distributed Real-Time Emulation of Formally-Defined Patterns for Safe 
Medical Device Control} \author{
Mu Sun \quad Jos\'e Meseguer
\institute{University of Illinois at Urbana-Champaign\\
Illinois, USA}
\email{musun(at)illinois.edu \quad meseguer(at)illinois.edu}
}
\def\titlerunning{Distributed Real-Time Emulation of Formally-Defined Patterns for Medical Device Control}
\def\authorrunning{M. Sun \& J. Meseguer}

\begin{document}

\maketitle

\input{abstract}
\input{introduction}
\input{overview}
\input{background}
\input{io-spec}
\input{time-spec}
\input{case-study}
\input{issues}
\input{conclusion}

{\bf Acknowledgments}
We would like to thank all the people of the UIUC Maude Seminar group and the 
the UIUC MD PnP group who have provided many valuable insights and suggestions 
towards improving this work. In particular we would like to thank Musab 
Al-Turki, who pioneered the effort in executing Maude models in physical time, 
for providing preliminary guidance on this work. This work was supported in 
part by ONR grant N000140810896 and by NSF grants 0720482 and 0834709.

\bibliographystyle{plain}
\bibliography{bibl}

\end{document}

%% file: abstract.tex
\begin{abstract}

Safety of medical devices and of their interoperation is an unresolved issue 
causing severe and sometimes deadly accidents for patients with shocking 
frequency.  Formal methods, particularly in support of highly reusable and 
provably safe patterns which can be instantiated to many device instances can 
help in this regard.  However, this still leaves open the issue of how to pass 
from their formal specifications in logical time to executable emulations that 
can interoperate in physical time with other devices and with simulations of 
patient and/or doctor behaviors.  This work presents a specification-based 
methodology in which virtual emulation environments can be easily developed 
from formal specifications in Real-Time Maude, and can support interactions 
with other real devices and with simulation models. This general methodology is 
explained in detail and is illustrated with two concrete scenarios which are 
both instances of a common safe formal pattern: one scenario involves the 
interaction of a provably safe pacemaker with a simulated heart; the other 
involves the interaction of a safe controller for patient-induced analgesia 
with a real syringe pump.

\end{abstract}

%% file: introduction.tex
\section{Introduction}

\label{sec:intro}

Each year, just in the US hospitals, a shocking and almost unacceptable number 
of medical accidents occur. In a 2009 study, reports estimate 40,000 instances 
of medical harm occur daily, and from the 2005 through 2007 period, at least 
92,882 deaths were potentially preventable \cite{HealthGrades2009}. Many of 
these accidents happen due to mistakes and failures in the 
\emph{interoperation} of medical devices. A modern hospital's operating room is 
in fact a quite complex distributed embedded system (DES) with many devices 
involved in either passively monitoring the patient state or actively 
performing different parts of a procedure. Both the safety of the individual 
devices and the safety of interoperation between devices (and between the 
patients and doctors) are of paramount importance. Presently, this safety is 
not adequately guaranteed.

When the reported accidents are analyzed, it becomes clear that many of them 
could and should have been avoided if the DES formed by the devices, the 
patient, and the doctors had been properly designed and analyzed, so that many 
unsafe interactions become \emph{impossible} by design.  The use of formal 
methods can clearly help in this respect, and promising research advances have 
already been made in this direction (see, e.g., 
\cite{alur04,arney07,ray04,jetley06}).

In our recent work \cite{tech-rep}\cite{sun-meseguer-sha-wrla10}, we have 
developed a scalable and highly reusable approach to the safety of medical 
devices by means of \emph{formally verified patterns} that: (i) are formally 
specified as real-time rewrite theories in Real-Time Maude; (ii) are generic, 
so that they apply not to a single device but to a wide range of devices, and 
are therefore specified as \emph{parameterized} modules; (iii) come with 
explicit \emph{formal requirements} (specified in their parameter theories) 
that must be met by any pattern instantiation to be correct; and (iv) come with 
\emph{formal safety guarantees} that will be satisfied by any correct 
instantiation of the pattern. For example, in \cite{tech-rep} we present one 
such pattern and show how it can be instantiated to obtain safe controllers for 
quite different devices, such as a pacemaker, an infusion pump for analgesia, 
and the interoperation of a ventilator with an X-ray machine.

However, there is still a substantial gap between the verified safety of 
designs in formal specifications and the actual safety of real medical devices, 
patients and doctors in an operating room for at least two reasons.  First, the 
formal specifications are somewhat idealized abstractions in which, for 
example, time is not the actual physical time that devices need to operate in, 
but logical time, and the code of the actual devices and controllers is not 
used but instead some formal specifications are used.  Second, it is important 
to consider not just the safety of a single device or small collection of 
devices, but also that of their \emph{interoperation} with other devices and 
with the patient and the doctors.  This work takes some first steps towards the 
goal of bridging the gap between formal specifications and actual devices in a 
hospital to help ensure that safety properties are preserved in the passage 
from specification to actual code and physical devices.  To achieve this goal
we propose the use of \emph{virtual emulation environments} in which:
\begin{enumerate}
\item formally verified patterns \cite{sun-meseguer-sha-wrla10}  can be 
instantiated to obtain various concrete specifications of desired devices and 
controllers;

\item the so-obtained formal executable specifications of devices and 
controllers are used \emph{directly} to generate emulators that perform the 
same specified behavior in physical time;

\item actual devices, as well as actual executable models of patient and doctor 
behavior, can be seamlessly integrated with specification-based emulators to 
validate the safety not just of individual devices but also of various 
DESs that are needed in practice in actual operating room conditions and 
scenarios.
\end{enumerate}
The advantage of point (1) is a great degree of reusability, and amortizing the 
formal verification effort across potentially many devices.  The advantage of 
(2) is that, since each specification-based emulator executes the exact same 
formal specification that has been proved safe for the given device, the safety 
of such an emulator is automatically guaranteed, and a path remains open to 
correctly generate actual code for it preserving such safety in an actual 
implementation. The advantage of point (3) is that system experimentation with 
physical time and actual devices becomes available from very early in the 
design process and are available afterwards along the entire development 
process: initially, only specifications may be emulated; at intermediate 
stages, both specifications and actual devices form the virtual emulation 
environment; and in the end the emulating environment seamlessly becomes an 
actual implementation.

Technically, the way such virtual emulation environments are obtained from 
formal specifications is by using a key idea first demonstrated by Musab 
Al-Turki to semi-automatically pass from a Real-Time Maude formal executable 
specification operating in \emph{logical time} to a corresponding 
\emph{physical emulation} of the same specification operating in \emph{physical 
time} and possibly interacting with other devices in a distributed way.  The 
key observations are: (i) in Real-Time Maude rewrite rules are either 0-time 
rules requiring no time, or time-advancing rules moving the entire system 
forward in logical time; (ii) time advancing rules (typically a single such 
rule) can be physically implemented by an external object that sends time ticks 
according to physical time;  (iii) although so-called 0-time rules do take some 
physical time to be executed, if this time is small enough in comparison with 
the time granularity of the physical time period chosen, for all practical 
purposes they can be considered to take 0 time units to execute; and (iv) the 
Maude infrastructure for Maude computations to interact with external objects 
via sockets can be used to interface the Maude objects in the formal 
specification with the external ticker object and also to other external 
devices.

\subsection{Our Contribution.}  This is the first work we are aware of in which 
formal specifications of real-time components are directly used in the area of 
distributed embedded systems for medical applications to obtain a virtual 
emulation environment in which specifications, patients, doctors, and actual 
devices can be emulated in physical time, and such that the correctness of 
verified specifications is preserved, provided adequate timing 
constraints are obeyed (see Section \ref{sec:issues}).  This work is a first 
step towards a seamless integration of formal specification, verification, and 
system development and testing for safe medical systems.  The associated notion 
of a virtual emulation environment plays a crucial role in passing from 
specifications to code and devices, and from logical time to physical time.  We 
have demonstrated both the feasibility and the usefulness of these methods in 
two concrete scenarios: one in which a pacemaker interacts adaptively in 
physical time with a simulated model of a patient heart and keeps heart rates 
within a safe envelope; and another in which a safety controller for 
patient-controlled analgesia interacts in physical time with an actual drug 
infusion device and with a simulation of patient behavior.

In the passage from formal real-time specifications to their corresponding 
emulators there are additional novel contributions that were required for this 
work, including: (i) advancing time by the maximum time elapsable as opposed to 
by a fixed ticking period; (ii) handling asynchronous interrupts in addition to 
synchronous communication; (iii) emulating the interaction of real medical 
devices, patient models, and formal specifications; and (iv) generating a timed 
wrapper for each component specification almost for free with deterministic 
Real-Time Maude specifications using both a time ticker and the computation of 
the maximum time elapsable for each time advance.

The paper is organized as follows: Section \ref{sec:overview} provides the high 
level ideas of the framework from formal patterns to real-time execution. 
Section \ref{sec:maude-background} covers the basics of Real-Time Maude and 
Maude's support for socket programming. Section \ref{sec:io} and 
\ref{sec:emulation} describes the core of the execution framework which allows 
seamless passage from Real-Time Maude specifications to execution with physical 
time and physical devices. Section \ref{sec:case} covers case studies for a 
pacemaker and a syringe pump to evaluate the feasibility of using formal models 
to execute medical devices. Finally, we describe some fundamental assumptions 
required for formal model execution to work for medical devices in Section 
\ref{sec:issues}, and we conclude in Section \ref{sec:conc}.

%% file: overview.tex
\section{Overview of the Model Execution Framework}

\label{sec:overview}

\begin{figure}
\centering \includegraphics[width=9cm, angle=0]{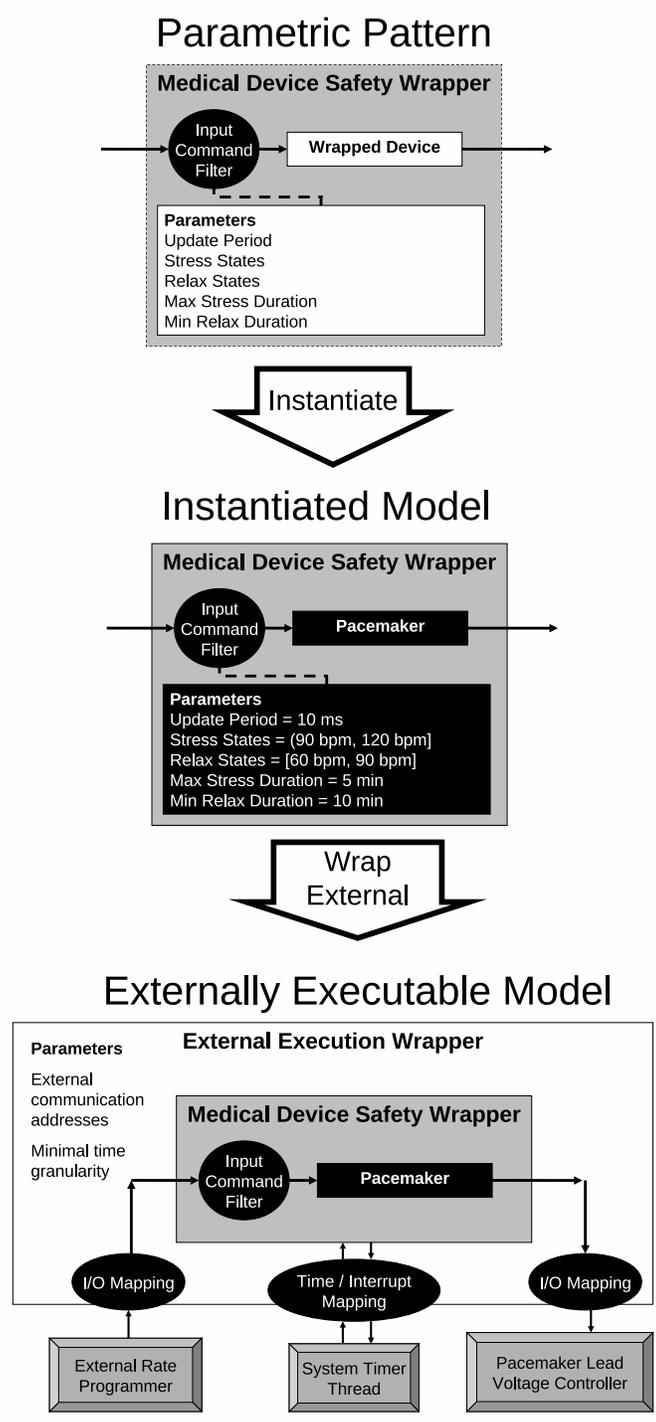}
\caption{From Formal Patterns to Real-Time Execution}
\label{fig:pattern-to-execution}
\end{figure}

An envisioned design framework from generic design patterns to executable 
specifications is shown in Figure \ref{fig:pattern-to-execution}. We start with 
a safety pattern, which is a parameteric module with well-defined parameters 
with formal requirements provided by an input theory. The next stage is to 
instantiate the pattern to a concrete instance, which will of course still 
satisfy all the safety properties ensured by the pattern. Finally, in the last 
stage (the focus of this paper), the entire executable specification of a 
system can really be executed in the real world by encapsulating the 
specification in an external wrapper for model execution. Figure 
\ref{fig:pattern-to-execution} illustrates these various stages for the design 
of a cardiac pacemaker system.

The first step formally defines a safety pattern as a parameterized module. In 
our pacemaker example, the pattern is a generic safety wrapper for medical 
devices. We briefly describe the pattern in this paper, summarizing the details 
presented in \cite{tech-rep}\cite{sun-meseguer-sha-wrla10}. The safety wrapper 
filters the input commands, so that state changes in a medical device fall 
within safe physiological ranges and constraints. The white boxes in the 
diagram denote pattern parameters that must be instantiated. For the device 
safety wrapper these parameters include the type of device that is being 
considered, the period for updating device states, states of the device 
considered stressful for the patient, etc. Aside from the parameters, there are 
also formal constraints that these parameters must conform to in order for a 
parameter instance to be acceptable. For example, stress states and relaxed 
states must be disjoint for a device.

The next stage is to instantiate the parameters of the pattern. In our 
example, the medical device safety wrapper is instantiated to filter the input 
commands to a pacemaker. The state of the pacemaker is assumed to be its pacing 
rate. The necessary parameters are then filled in. The period for updating the 
pacing rate is every 10 ms; pacing rates between 90 bpm to 120 bpm are 
considered stressful for the patient; etc. This instantiated model will satisfy 
the safety properties guaranteed by the safety pattern, provided all the 
parameters satisfy the necessary constraints and formal parameter requirements. 
Furthermore, once instantiated, the wrapped pacemaker is just another 
executable Real-Time Maude model. Thus, we can use this model for simulation and
model checking purposes in Real-Time Maude.

\begin{figure}
\centering \includegraphics[width=6cm, angle=-90]{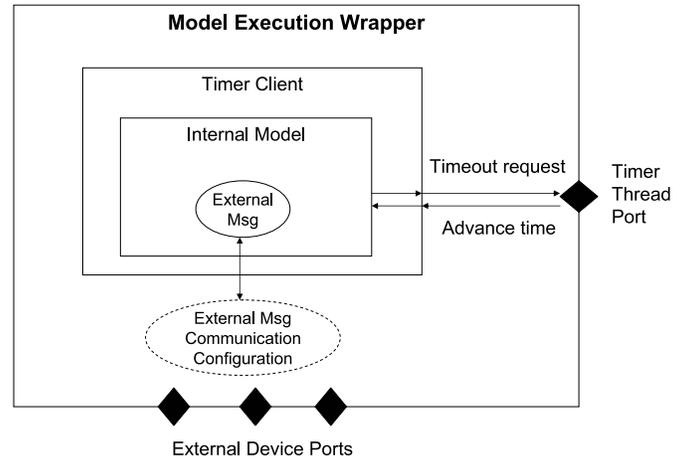}
\caption{Real-Time Model Execution Wrapper}
\label{fig:mdewrapper}
\end{figure}

In the final stage, which is the main focus of this paper, we transform the 
model to execute in {\em real world} time with physical devices in a medical 
device emulation environment. For this purpose, we take the model of the 
wrapped pacemaker and wrap it again in an external execution wrapper (Figure 
\ref{fig:mdewrapper}). The execution wrapper is responsible for conveying to 
the model the notion of real world time as well as providing a communication 
interface to the external world. A dedicated timer thread is responsible for 
``ticking'' the model by sending a minimal number of messages to advance the 
model's logical time. The timer thread also intercepts all asynchronous 
(interrupt) messages and relays them to the model. Another aspect of the 
execution wrapper is the ability to map external I/O messages to communicate 
with the external devices. For example, in the pacemaker specification, an 
internal message called {\em paceVentricle} may be mapped into an entire client 
configuration to send a message for setting the final voltage on a pacing lead.

For the design of a medical device or, more generally, of any safety-critical 
system, all the stages can work together to achieve a modular component design, 
and to support experimentation and testing in the context of other real devices 
that the safe component being designed has to interact with. In the first 
stages of safety pattern specification, we create a parameteric formal 
specification that essentially isolates important safety properties of a device 
from the rest of the system. We then can use theorem proving techniques to 
provide provable properties of the safety pattern. Although theorem proving 
may be time-consuming, as we show with our case studies one safety pattern can 
be applied to many different applications, so the time spent proving properties 
of the pattern is well worth the effort. The second stage with pattern 
instantiation is necessary to obtain a fully specified and executable model. 
The last stage of course takes the existing models, with minimal auxiliary 
information for external interfaces, and provides an executable prototype 
essentially for free. In this way, it becomes possible to emulate the behavior
of safe medical components in an experimental environment involving interactions
with real medical devices.

%% file: background.tex
\section{Real-Time Maude and Sockets}
\label{sec:maude-background}

In this section we briefly cover the important constructs we used from 
Real-Time Maude and Maude Sockets. We assume the reader is familiar with basic 
Maude constructs including modules ({\tt mod}), sorts ({\tt sort}), operators 
({\tt op}), unconditional and conditional equations ({\tt eq} and {\tt ceq}) 
and unconditional and conditional rules ({\tt rl} and {\tt crl}).

\subsection{Full Maude and Real-Time Maude}

Full Maude \cite{fullmaude} is a Maude interpreter written in Maude, which in
addition to the Core Maude constructs provides syntactic constructs such as
object oriented modules. Object oriented (OO) modules implicitly add sorts
{\tt Object} and {\tt Msg}. Furthermore, OO-modules add a sort called {\tt
Configuration} which consists of a multiset of terms of sort {\tt Object} or {\tt
Msg}.

Objects are represented as records:

\begin{footnotesize}
\begin{alltt}
< {\em objectID} : {\em classID} | {\em AttributeName} : {\em Attribute}, ... >
\end{alltt}
\end{footnotesize}

Rewrite rules are then used to describe state transitions of objects
based on consumption of messages. For example, the following rule expresses
the fact that a pacemaker object consumes a message to set the pacing
period to T:

\begin{footnotesize}
\begin{alltt}
rl setPeriod(pm, T)
   < pm : Pacing-Module | pacing-period : PERIOD >
   => < pm : Pacing-Module | pacing-period : T > .
\end{alltt}
\end{footnotesize}

Real-Time Maude \cite{rt-maude} is a real-time extension of 
Maude in Full Maude. It adds syntactic constructs for defining timed 
modules. Timed modules automatically import the {\tt TIME} module, which defines 
the sort {\tt Time} (which can be chosen to be discrete or continuous) along
with various arithmetic and comparison operations on {\tt Time}. Timed modules
also provide a sort {\tt System} which encapsulates a {\tt Configuration} and
implicitly associates with it a time stamp of sort {\tt Time}. After defining a
time-advancing strategy, Real-Time Maude provides timed execution ({\tt
trew}), timed search ({\tt tsearch}), which performs search on
a term of sort {\tt System} based on the time advancement strategy, and
timed and untimed LTL model checking commands.

\subsection{Deterministic Timed Rewriting in Real-Time Maude}

We are interested in emulations of real-time systems specified in Real-Time 
Maude. For useful real execution, a self-evident condition is that the 
time-advancing rewrite rules in the specification should be deterministic. This 
can be achieved by defining only one time-advancing rewrite rule on the system 
with two auxiliary operators {\em tick} and {\em mte} \cite{rtmaude-manual}. In 
our specification this is captured in {\tt TIME-ADV-SEMANTICS}, which is 
included by all other timed modules:

\begin{footnotesize}
\begin{verbatim} 
(mod TIME-ADV-SEMANTICS is ...
  op mte : Configuration ~> TimeInf . ...
   eq mte(none) = INF .
   eq mte(C C') = minimum(mte(C), mte(C')) .
  op tick : Configuration Time ~> Configuration . ...
   eq tick(none, T) = none .
   eq tick(C C', T) = tick(C, T) tick(C', T) .
  op def-te : -> Time .
  op max-te : -> TimeInf .

   crl {GC} => {tick(GC, T)} in time T if T <= mte(GC) [nonexec] .
endm)
\end{verbatim}
\end{footnotesize}

The rewrite rule at the end is assumed to be the only timed rewrite rule (a 
rewrite rule that advances the time stamp of the system) in the system 
specification. The {\tt mte} operator defines the maximum time elapse before 
any 0-time rewrite rule can be applied. The {\tt tick} operator defines how 
the system state changes due to time advancement between applications of 
0-time rewrite rules. We also define a default time elapse, {\tt def-te}, 
and maximum time elapse, {\tt max-te}, inside the module to be used as 
parameters during real execution.

\subsection{Socket Programming in Maude}

Maude supports the Berkley sockets API for TCP communication. This is done by 
having a special gateway object, denoted {\tt <>}, to consume all the messages 
responsible for setting up sockets and communicating to an external environment 
(e.g. {\tt createClientTcpSocket}, {\tt send}, {\tt receive}). The gateway 
object will also generate messages upon status updates from the socket (e.g. 
{\tt sent}, {\tt received}, {\tt closedSocket}). Consuming and generating 
messages from the gateway object is captured by external rewrite rules which 
can be executed using the {\tt erew} command in Core Maude. An important thing 
worth pointing out about external rewrite rules is that {\em external rewrite 
rules are only applied when no internal rewrite rules can be applied}. Also, 
using external rewrite rules with Real-Time Maude specifications (built on top 
of Full Maude) requires reflecting the specification down to a Core Maude 
module before executing.

%% file: io-spec.tex
\section{Mapping Internal Messages to External I/O}
\label{sec:io}

Validating the design of a device in an execution environment requires handling 
its outputs. After all, the end validation of a system's behavior is based on 
its outputs. Thus, it seems reasonable to talk about how internal messages in 
the model can be converted into messages for communicating with the external 
world. This section also serves as an explanation for unfamiliar readers of how 
Maude sockets are used.

In order to talk about external communication, we must first define in the 
model what is external. The model will have an internal distributed actor 
configuration with internal messages as well as messages to be output to the 
external world. Thus, the first definition is {\tt EXTERNAL-CONFIGURATION} 
which defines external messages {\tt ExtMsg} as subsort of {\tt Msg}. 
Furthermore, external messages are classified in terms of incoming external 
messages {\tt InExtMsg} and outgoing external messages {\tt OutExtMsg}. A 
configuration is called {\em open} if there are external messages present in 
the configuration: either an incoming external message has not been delivered, 
or an outgoing external message has not been sent. The predicate {\tt open?} is 
defined accordingly.

\begin{footnotesize}
\begin{verbatim}   
subsorts InExtMsg OutExtMsg < ExtMsg < Msg .
op open? : Configuration -> Bool .
 eq open?(C C') = open?(C) or open?(C') .
 eq open?(O) = false .
 eq open?(M) = M :: ExtMsg .
\end{verbatim}
\end{footnotesize}

Actually sending an external message may be more complex than just forwarding 
the message through the gateway object. External messages may not be the same 
in the internal configuration and in the external configuration. For example, a 
simple output message in the internal configuration may need to be mapped to a 
client object that initiates the communication to deliver the message. 
Operators {\tt in-adapter} and {\tt out-adapter} are defined to perform these 
mappings from external message client configurations to internal messages.

An example of an output adapter for a pacemaker message to beat the heart
may be:

\begin{footnotesize}
\begin{verbatim}
eq out-adapter(shock)
    = createSendReceiveClient(pacer-client, "localhost", 4451, "SetLeadVoltage 5V")
\end{verbatim}
\end{footnotesize}

In this example, the message {\tt shock} is transformed into a client object 
which sends a message on port 4451 with the string {\tt "SetLeadVoltage 5V"} 
indicating that the proxy server will then proceed to set a 5V voltage on the 
pacemaker lead.

\subsection{One-Round Communication Clients}

Once the external message is mapped into a client configuration, we must define
the rewrite rules to specify how the communication protocol works with the
external device. Here we describe a simple {\tt SEND-RECEIVE-CLIENT} which is
responsible for establishing communication, sending a message, receiving a
reply, and then closing the communication. Although simple, this type of
protocol is sufficient for most of the communication for medical devices we have
used in our case studies.

\begin{footnotesize}
\begin{verbatim}
(mod SEND-RECEIVE-CLIENT is ...
  op createSendReceiveClient : Oid String Nat String -> Configuration .
   eq createSendReceiveClient(CLIENT, ADDRESS, PORT, SEND-CONTENTS)
      = < CLIENT : SendReceiveClient | ... >
      createClientTcpSocket(socketManager, CLIENT, ADDRESS, PORT) .
  op msg-received : Oid String -> InExtMsg .
...endm)
\end{verbatim}
\end{footnotesize}

After creating the client and establishing communication, the client goes into 
one round of send and receive before the socket is closed. Once the socket is 
closed, the entire client object is converted into one reply message to be 
delivered to the internal configuration using the operator {\tt msg-received}.

\begin{footnotesize}
\begin{verbatim}
--- send contents
rl createdSocket(CLIENT, socketManager, SOCKET-DST)
   < CLIENT : SendReceiveClient | ... send-contents : SEND-CONTENTS >
   => < CLIENT : SendReceiveClient | ... > send(SOCKET-DST, CLIENT, SEND-CONTENTS) .
--- receive contents
rl sent(CLIENT, SOCKET-DST) < CLIENT : SendReceiveClient | ... >
   => < CLIENT : SendReceiveClient | ... > receive(SOCKET-DST, CLIENT) .
--- close socket
rl received(CLIENT, SOCKET-DST, RECEIVE-CONTENTS) < CLIENT : SendReceiveClient | ... >
   => < CLIENT : SendReceiveClient | ... recv-contents : RECEIVE-CONTENTS >
      closeSocket(SOCKET-DST, CLIENT) .
--- done
rl closedSocket(CLIENT, SOCKET-DST, "")
   < CLIENT : SendReceiveClient | ... recv-contents : RECEIVE-CONTENTS >
   => msg-received(CLIENT, RECEIVE-CONTENTS) .
\end{verbatim}
\end{footnotesize}

%% file: time-spec.tex
\section{Distributed Emulation of Safe Medical Devices}
\label{sec:emulation}

The external execution wrapper is an object that encapsulates the original 
formal model. It is primarily responsible for interfacing constructs between 
the physical world (the real interfaces to devices) and the logical world (the 
world as seen by the formal model). In particular, the execution wrapper is 
responsible for conveying the measurement of real time elapsed to the model and 
also for mapping logical communication messages to communication configurations 
that can deliver the message to real devices. The most important feature of the 
external execution wrapper is its modularity. Aside from adding the minimal 
information about how to map external I/O to messages in the model, no further 
specifications are required to execute the logical model within an external 
environment.

\subsection{Mapping Logical Time to Physical Time}

\begin{figure}
\centering \includegraphics[width=14cm, angle=0]{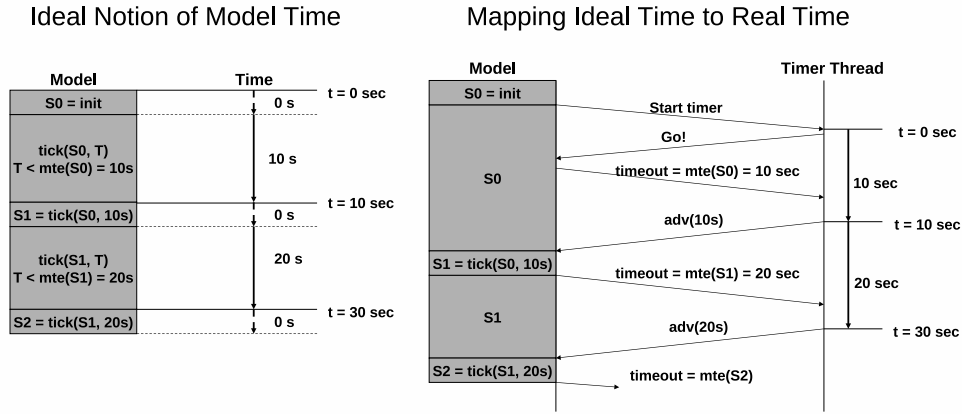}
\caption{From Ideal Time Advancement Semantics to Physical Time Advancement}
\label{fig:time}
\end{figure}

As mentioned earlier, time advancement of the system is achieved by defining 
the {\em tick} and {\em mte} operators. Ideally the system continuously evolves 
over time (possibly nondeterministically). Of course, we cannot capture the 
notion of continuous time without abstractions in the model, so to advance time 
discretely, an {\em mte} (maximum time elapsable) operator is introduced. A 
correctly defined {\em mte} operator ensures that if a system is in state $S$, 
then for any time $T < mte(S)$, no 0-time rewrite rules (state transitions)
 can apply to $tick(S, T)$. That is, if a system is in state $S$, and $T \leq 
mte(S)$, then $tick(S, T)$ will be equivalent to the state $S$ advancing in 
continuous time for $T$ time units. This ideal semantics of time is shown on the 
left side of Figure \ref{fig:time}. The figure shows that 0-time rewrite 
rules are assumed to take zero time, and ideally, the system continuously 
evolves over time between the 0-time rewrite rules.

Of course, in a real execution of the model, the ideal notion of time with 
0-time rewrite rules and time-advancing rules is only an idealized 
abstraction. Performing rewrites cannot take zero time, and we cannot 
continuously rewrite states of the system over time. We could of course create 
a model in discrete time with very fine time granularity and drive it by a high 
frequency clock like in hardware. However, this would introduce a lot of 
unnecessary overhead in terms of communication of timing messages and 
performing rewrite rules to change the model for every clock tick. We resolve 
this problem by observing that the actual internal state of the model is not 
important at most instants in time unless it is communicating with the external 
world. The model states only generate output messages with 0-time rewrite 
rules, so we can essentially let the model in state $S$ remain unaffected by 
the passage of time until the next time instant in which a 0-time rewrite 
rule can be applied; this is exactly $mte(S)$ time units later. This method of 
driving execution is shown in the right part of Figure \ref{fig:time}. We have 
created a dedicated timer server thread (in Java) that has access to the system 
time. When the execution of the wrapped model starts, it will send a start 
request which includes the time units of the model or the minimum granularity 
of time for model execution in milliseconds. Once the timer thread processes 
all the initial information, it will send a {\em Go!} message to signal the 
model to start executing. The model then calculates the maximum time elapsable 
(which is 10 seconds in the example) and sends this information to the timer 
thread. The model then proceeds to sleep until the timer thread wakes it in 
time for the next 0-time rewrite rule. The process then continues. There are 
two key points to notice about this example. The input and output messages from 
the model may be delayed by an amount of time equal to the communication jitter 
plus the time to complete rewriting. Normally this delay is on the order of 10 
ms, but this is still suitable for medical devices which normally receive 
commands on the order of seconds or more. Also, the timer thread sets the 
timeout from the last time it sent a time advancement message to the model and 
not from the time it receives the $mte$ message from the model. This ensures 
that clock skew and jitter are bounded over time.

\subsection{Synchronous Timed Execution}

The {\em communication wrapper} ({\tt commwrap}) is represented as an object 
with the attributes for the communication state, the socket information for 
communication, and the internal wrapped (Real-Time Maude) system model being 
executed. The top level system is of sort {\tt CommWrapConfiguration} for any 
communicating model.

\begin{footnotesize}
\begin{verbatim}
op commwrap : Configuration -> CommWrapConfiguration .
op wrap-client : Configuration -> Configuration .
 eq wrap-client(C) = < client : TickClient |
                       state : start, internal : [ {C} in time 0 ], socket-name : no-oid > .
op init-client : -> CommWrapConfiguration .
 eq init-client = commwrap( <> wrap-client(internal)
                            createClientTcpSocket(socketManager, client, addr, port) ) .
\end{verbatim}
\end{footnotesize}

The communication wrapper initializes a wrapped communication client that 
receives messages from the tick server (a Java thread executing in real-time 
that sends it messages for time advancement). After creating the TCP socket, 
the first message sent from the client to the tick server is the 
time-granularity ({\tt time-grain}), which is a rational number specifying the 
number of milliseconds in one time unit. Then, the actual execution starts when 
the communication wrapper receives a {\tt GO} message from the tick server. The 
time when the tick server sends the {\tt GO} message is the starting point from 
which time elapses are being measured. Upon receiving the {\tt GO} message, the 
formal model will immediately start to execute ({\tt state : run}).

\begin{footnotesize}
\begin{verbatim}   
rl [send-init] :
 commwrap( <> createdSocket(...) < client : TickClient | ... > )
  => commwrap( <> < client : TickClient | ... > send(..., string(time-grain)) ) .
rl [wait-for-go] :
 commwrap( <> sent(...) < client : TickClient | ... > )
  => commwrap( <> < client : TickClient | ... > receive(...) ) .
rl [start-running] :
 commwrap( <> received(..., "GO\r\n") < client : TickClient | ... > )
  => commwrap( <> < client : TickClient | state : run, ... > .
\end{verbatim}
\end{footnotesize}

The formal model executes until {\tt mte} becomes non-zero (no other 0-time 
rewrite rules can be applied), and the model sends a message to request the 
next time advancement message after the maximum time elapse and blocks. After 
sending this waiting duration, the tick server will sleep for this time 
duration and then send a time advancement message when the time has expired. 
The model will then advance time (tick) the model for the time duration expired
and perform 0-time rewrite rules. The model now blocks again for the next {\tt
mte}, and the cycle repeats.

\begin{footnotesize}
\begin{verbatim}   
crl [request-wait-timer] :
 commwrap( <>
   < client : TickClient |
      state : run,
      internal : [ {C} in time T ], ... > )
 => commwrap( <>
   < client : TickClient |
      state : request, ... >
   send(..., string(mte(C, T))) )
 if mte(C,T) :: TimeInf /\ mte(C,T) > 0 /\ not open?(C) . ...

rl [block] :
 commwrap( <> sent(...)
   < client : TickClient |
      state : request, ... > )
 => commwrap( <>
   < client : TickClient |
     state : wait, ... >
     receive(SOCKET-NAME, client) ) .

rl [wake-up] :
 commwrap( <> received(..., ADV-STR)
   < client : TickClient |
      state : wait, ... > )
 => commwrap( <>
   < client : TickClient |
      state : run,
      internal : [ {tick(C, rat(ADV-STR))} in time rat(ADV-STR) in time T ], ... > ) .
\end{verbatim}
\end{footnotesize}

\subsection{Handling Asynchronous External Events}

So far, the model can only handle synchronous events (polling and blocking 
communication). However, in general a useful design must be able to react to 
external events from the environment. For example, an EKG sensor detects a QRS 
waveform, and sends this information to the pacemaker. This points to the fact 
that our model needs to be able to handle external events asynchronously.

\begin{figure}
\centering \includegraphics[width=8cm, angle=0]{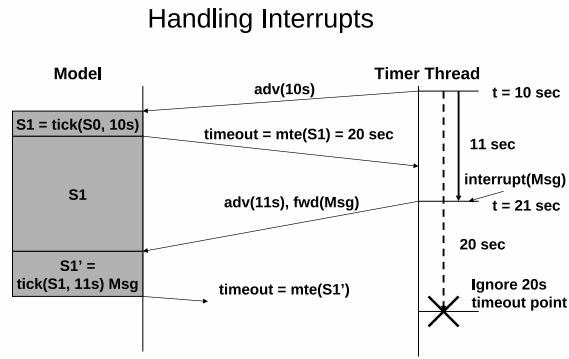}
\caption{Handling Interrupts and Asynchronous Communication Semantics}
\label{fig:async}
\end{figure}

An external message would trigger a 0-time rewrite rule to receive the message 
by some object and process it. More precisely, if we have $C_M$ and $C_{Ext}$ 
as the model configuration and the external (environment) configuration 
respectively, the maximal time elapse for the system $C_M C_{Ext}$ should be 
$min(mte(C_M), mte(C_{Ext}))$, where $mte(C_{Ext})$ denotes time duration 
before the next interrupt message. This semantics is captured by having 
interrupt messages forwarded by the timer thread, as shown in Figure 
\ref{fig:async}. The timer thread will only check for interrupts when it is 
waiting for the next timeout, so when the interrupt message arrives, it will 
wake up and immediately forward the interrupt message to the model with the 
amount of time that has elapsed. Any future timeouts are canceled. Introducing 
the notion of interrupts requires us to modify the wake-up rule for the model 
to not only advance time, but also check for potential interrupt messages as 
well.

\begin{footnotesize}
\begin{verbatim}
rl [wake-up] :
commwrap( <> received(client, SOCKET-NAME, INTR-STR)
  < client : TickClient |
  state : wait,
  internal : [ {C} in time T ],
  socket-name : SOCKET-NAME > )
=>
commwrap( <> < client : TickClient |
  state : run,
  internal : [
    {tick(C, recv->rat(INTR-STR)) recv->conf(INTR-STR)}
      in time recv->rat(INTR-STR) in time T
  ],
  socket-name : SOCKET-NAME > ) .
\end{verbatim}
\end{footnotesize}

%% file: case-study.tex
\section{Case Studies}

\label{sec:case}

\subsection{A Pattern for Medical Device Execution}

\begin{figure}
\centering \includegraphics[width=10cm, angle=0]{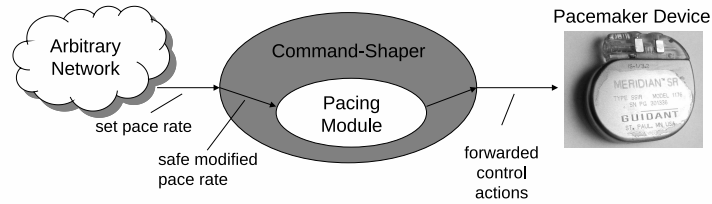}
\caption{Command Shaper Pattern for a Pacemaker}
\label{fig:commandshaper}
\end{figure}

We briefly described in the introduction at a high level that the model 
execution framework is to support rapid prototyping of instantiated medical 
device safety patterns. In \cite{sun-meseguer-sha-wrla10} and \cite{tech-rep}, 
we have described in detail the command shaper pattern for medical device 
safety. In essence, the command shaper pattern can modify commands to an 
existing medical device to guarantee specific safety properties in terms of 
limiting durations of stressful states and limiting the rate of change (Figure 
\ref{fig:commandshaper}).

\subsection{Pacemaker Simulation Case Study}

One of the applications for the command shaper pattern is a pacemaker system 
\cite{sun-meseguer-sha-wrla10}. At a high level the safety properties 
guaranteed by the command shaper pattern is that the pacemaker will not pace at 
fast heart rates too frequently or for too long, and the pacing rate will 
change gradually. We omit the details of instantiating the medical device 
pattern, but the final wrapper object provided by the pattern is:

\begin{footnotesize}
\begin{verbatim}
(tomod PARAM-PACEMAKER is pr EPR-WRAPPER-EXEC{Safe-Pacer} ...
   eq wrapper-init =
      < pacing-module : EPR-Wrapper{Safe-Pacer} |
         inside :
            < pacing-module : Pacing-Module |
               nextPace : t(0),
               period : safe-dur >,
         val : safe-dur,
         next-val : safe-dur,
         disp : t(period),
         stress-intervals : (nil).Event-Log{Stress-Relax} > .
... endtom)
\end{verbatim}
\end{footnotesize} 

This says that a wrapper is placed around a pacing module, and the initial 
pacing rate is set as the default safe-duration ({\tt safe-dur} is 750 ms or 80 
heart beats per minute). Verifying this instantiation (with a simple pacemaker 
lead model \cite{tech-rep}) indicates that the safety properties are met by the 
pattern. However, with the power of the model emulation framework, we can 
immediately use this specification to run with an actual pacemaker. In this 
paper we demonstrate this emulation capability not on an actual pacemaker but 
on a pacemaker simulator (a Java widget that receives messages about when to 
pace and draws a simple line graph resembling an ECG trace). Before the system 
can be emulated with the pacemaker simulator, some interface information must 
be provided. The entire module providing all the necessary interface 
information is shown below:

\begin{footnotesize}
\begin{verbatim}
(mod CREATE-TICKER is
   inc PARAM-PACEMAKER .
   inc TIME-CLIENT .
   inc SEND-RECEIVE-CLIENT .

   eq addr = "localhost" .
   eq port = 4444 .

   eq def-te = 1 .
   eq max-te = INF .
   eq time-grain = 10 . --- milliseconds

   op pacer-client : -> Oid .
   eq internal = wrapper-init .
   
   eq out-adapter(shock)
      = createSendReceiveClient(pacer-client, "localhost", 4451, "shock") .
   eq in-adapter(msg-received(pacer-client, "shocked\n"))
      = set-period(pacing-module, 50) .
endm)
\end{verbatim}
\end{footnotesize} 

The module first indicates that the TCP
socket interface to the pacemaker simulator is {\em localhost} on {\em port}
4444. The default time elapse for one tick is 1 time unit. The maximum time elapse
for one tick step is infinity (i.e. there is no maximum). The duration of one time
unit is 10 milliseconds. The time units are in terms of milliseconds since the
minimum time granularity provided by the Java time interfaces is 1 millisecond.

The equation for {\tt internal} specifies that the internal configuration to be
executed is the configuration defined by {\tt wrapper-init} (as defined in {\tt
PARAM-PACEMAKER}). Also, the last two equations specify that the output message
shock should be mapped to a string ``shock'' sent over the socket, and upon
receiving the acknowledgment message ``shocked'' set the pacing period to 500
ms (120 bpm - a really fast heart rate). The last equation creates the scenario
where a stressful heart rate is always being sent to the pacing module. Since
the command shaper pattern should prevent this unsafe behavior, we should see
the pacing automatically slow down from 120 bpm after some time interval.

\begin{figure}
\centering \includegraphics[width=6cm, angle=0]{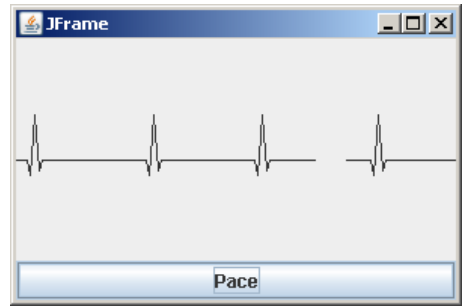}
\caption{Trace from Pacemaker Simulator}
\label{fig:pacing}
\end{figure}

\begin{figure}
\centering \includegraphics[width=12cm, angle=0]{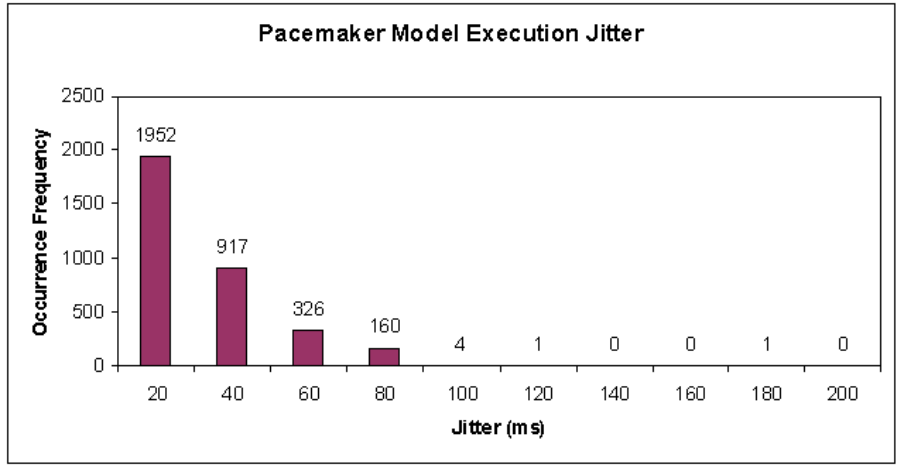}
\caption{Model Execution Jitter Distribution}
\label{fig:jitter}
\end{figure}

\begin{figure}
\centering \includegraphics[width=12cm, angle=0]{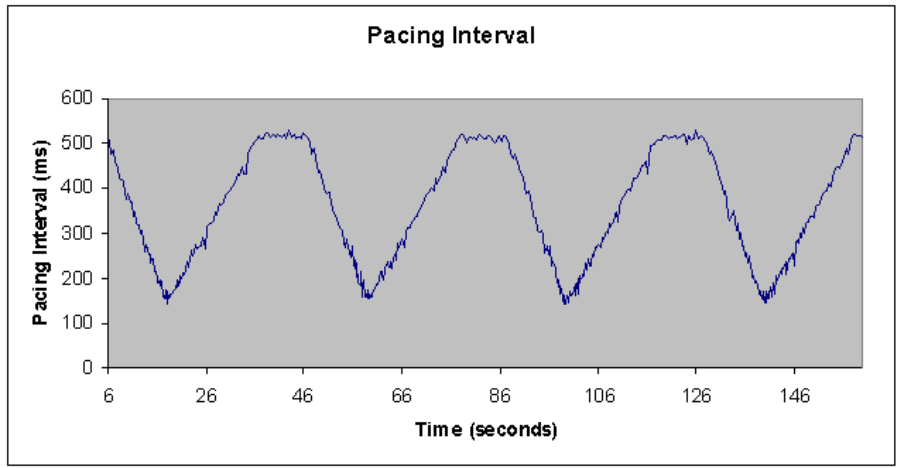}
\caption{Pacing periods recorded by the pacemaker simulator (jitter effects
are reflected by noise on the curve)}
\label{fig:pacingintervals}
\end{figure}

The module is executed by first reflecting the {\tt CREATE-TICKER} module down 
to Core-Maude (with the command {\tt show all CREATE-TICKER}), and executing 
with the {\tt erew} command. A snapshot of the ``ECG'' trace of the pacemaker 
simulator is show in Figure \ref{fig:pacing}. For validation, we measured the 
jitter for executing such a system -- the physical time required to completely 
execute 0-time rules and finish communication (Figure \ref{fig:jitter}). The 
results were obtained from a 1.67 GHz Dual-Core Intel Centrino with Maude 
running in Windows through Cygwin (tracing was turned off). The main thing to 
notice is that the jitter is mostly below 0.1 seconds and almost never exceeds 
0.2 seconds. This amount of jitter is tolerable since most medical devices need 
to respond in the order of seconds. The pacemaker is a bit more strict in terms 
of its timing requirements. To evaluate suitability for the pacemaker, we 
plotted the recorded the physical time duration between pacing events (Figure 
\ref{fig:pacingintervals}). Notice in this example the heart rate increases 
(duration decreases) up to a limit and then the heart rate starts to decrease 
(duration increases) and the cycle repeats. It is clear that the jitter in 
control seems tolerable since there are no sharp spikes in the graph of the 
pacing durations.

% The main thing to notice is that the pacing events (spikes in the graph) are 
% spaced relatively evenly apart. This shows that although the timed emulation of 
% these systems incurs rewriting delays for the zero timed rules, for the time 
% granularities operated by medical devices (on the order of 0.1 seconds), these 
% delays and jitters are immaterial.

\subsection{Syringe Pump Case Study}

The pacemaker emulation example was demonstrated through a simulated
pacemaker mostly because current pacemakers do not have external interfaces for
setting when to pace (and rightly so). However, for devices such as electronic
syringe pumps these interfaces are available. Syringe pumps and infusion pumps
in general deliver intravenous injections into a patient. For this scenario, we
assume that the syringe pump is delivering an analgesic (e.g.\ morphine) to the
patient, and we would like to prevent overdose. We assume that for a normal
patient overdoses do not occur at the base rate of infusion and can only result
if a bolus dose is administered too often for the patient. We again use the
command shaper pattern to limit the frequency and duration of bolus doses. The
instantiated pump is as follows:

\begin{footnotesize}
\begin{verbatim}
(tomod PARAM-PUMP is
   pr EPR-WRAPPER-EXEC{Safe-Pump} .
   pr DELAY-MSG .
...
   eq msgs-init =
      delay(set-mode(pump-module, bolus), t(9))
      delay(set-mode(pump-module, bolus), t(11))
      delay(set-mode(pump-module, bolus), t(12)) ... .
   eq wrapper-init =
      < pump-module : EPR-Wrapper{Safe-Pump} |
         inside :
            < pump-module : Pump-Module |
               mode : base
            > base,
         val : base,
         next-val : base,
         disp : t(period),
         stress-intervals : (nil).Stress-Relax-Log > .
... endtom)
\end{verbatim}
\end{footnotesize}

This module shows the initialized wrapper object for the pump, with the initial
state being the base rate of infusion. Furthermore, there is also a set of
delayed messages that will be sent to the pump. In the term {\tt msgs-init}, the
model will send bolus requests at 9 time units, 11 time units, 12 time units,
\ldots after the start of execution for the system. Again, creating a simulated
patient model, we can verify the safety of the instantiated pattern
\cite{tech-rep}. Instantiating the pump is similar to instantiating the
pacemaker, except that there are a few more types of output messages.

\begin{footnotesize}
\begin{verbatim}
(mod CREATE-TICKER is
   inc PARAM-PUMP .
   inc TIME-CLIENT .
   inc SEND-RECEIVE-CLIENT .

   eq addr = "localhost" .
   eq port = 4444 .

   eq def-te = 1 .
   eq max-te = INF .
   eq time-grain = 1000 . --- milliseconds

   ops pump-client pump-client' : -> Oid .
   eq internal = wrapper-init msgs-init .
   
   eq out-adapter(stop)
      = createSendReceiveClient(pump-client, "localhost", 1234, "STP") .
   eq out-adapter(base)
      = createSendReceiveClient(pump-client, "localhost", 1234, "RAT1")
      createSendReceiveClient(pump-client', "localhost", 1234, "RUN") .
   eq out-adapter(bolus)
      = createSendReceiveClient(pump-client, "localhost", 1234, "RAT2")
      createSendReceiveClient(pump-client', "localhost", 1234, "RUN") .
   var S : String .
   eq in-adapter(msg-received(pump-client, S))
      = none .
   eq in-adapter(msg-received(pump-client', S))
      = none .
endm)
\end{verbatim}
\end{footnotesize}

\begin{figure}
\centering \includegraphics[width=8cm, angle=0]{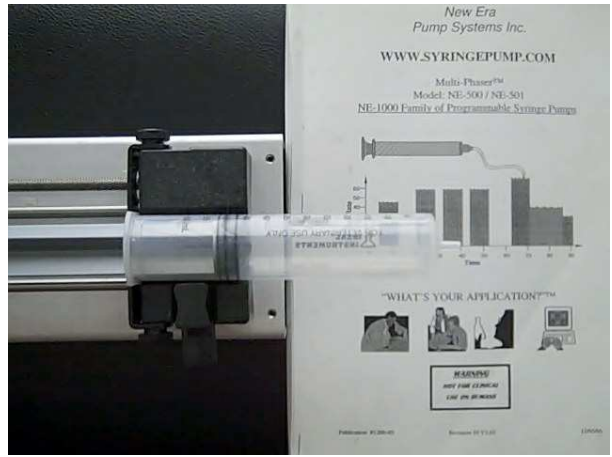}
\caption{Multi-Phaser NE-500 Syringe Pump}
\label{fig:pump}
\end{figure}

The model is communicating with {\em localhost} on {\em port} 4444. The time 
granularity is 1 second. The internal configuration being executed is the 
wrapped pump as well as the set of messages that will deliver bolus requests. 
The output requests are handled by a Java thread listening on port 1234 and 
forwarding the request string to the actual {\em Multi-Phaser NE-500} Syringe 
Pump (Figure \ref{fig:pump}). A few important requests to the pump are: {\tt 
STP} stop the pump, {\tt RAT <n>} set infusion rate to {\tt n} ml/hr, {\tt RUN} 
start the infusion. Reflecting down the {\tt CREATE-TICKER} module and 
executing with {\tt erew} will now control the physical pump motor!

\begin{figure}
\centering \includegraphics[width=8cm, angle=0]{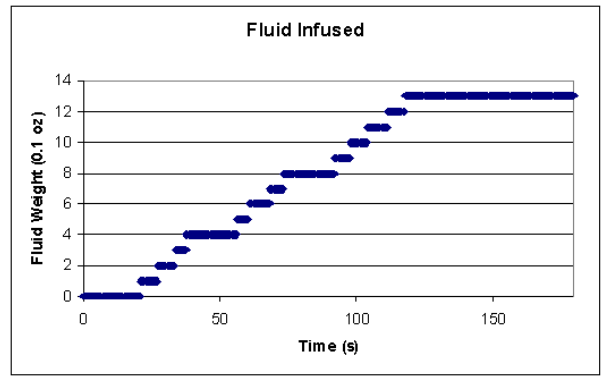}
\caption{Infusion Volume over Time}
\label{fig:pumpdata}
\end{figure}

As a validation for correct pump control, we used a Salter Brecknell 7010SB 
scale to weigh the amount of liquid infused from the syringe pump over time 
(Figure \ref{fig:pumpdata}). The data granularity is a bit rough since the 
scale can only measure within a precision of 0.1 oz. For this example, to 
clearly distinguish between two pump states, we let the base rate of infusion 
be zero (horizontal parts of the graph) and the bolus rate be the maximum 
infusion rate provided by the pump (positive sloped parts of the graph). Bolus 
requests are continuously sent to the pump. The safety properties require that 
bolus doses last no longer than 30 seconds, and there must be 10 seconds 
between bolus doses, and at most 3 bolus doses for a window size of 3 minutes. 
The graph validates that these properties are indeed satisfied for this 
particular execution of the pump.

%[perhaps some more validation results]

%% file: issues.tex
\section{Assumptions and Issues}

\label{sec:issues}

In this section we discuss the timing assumptions that need to be taken into
account to ensure that the emulation of a Real-Time Maude specification
correctly implements the logical time behavior.

\begin{figure}
\centering \includegraphics[width=10cm, angle=0]{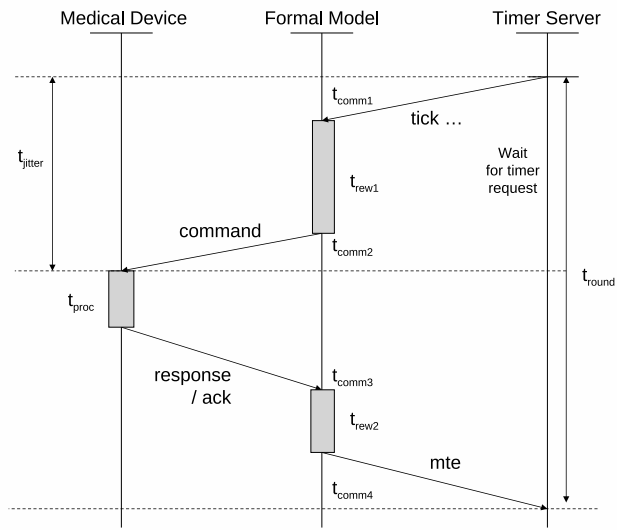}
\caption{Timing Considerations}
\label{fig:timing}
\end{figure}

Figure \ref{fig:timing} shows one round of communication between the time server
and the formal model.
$t_{comm,i}$ denotes the delays due to each stage of communication.
$t_{rew,i}$ denotes the delays incurred by each stage of rewriting. $t_{proc}$ 
denotes the time needed at the physical device interface to process
the commands. Thus, the entire time to finish a round is $t_{round} =
t_{comm1} + t_{rew1} + t_{comm2} + t_{proc} + t_{comm3} +
t_{rew2} + t_{comm4}$.

In logical time, no time advancement should actually take place in a 
communication round. All computation and message communication is assumed to 
take zero time. Of course, for proper timed operation we can relax these 
constraints to first allow the model to have a non-zero (but bounded) delay for 
these computations and communications. The maximum bound on these 
communications is $t_{round} \leq mte(C_{next})$. Otherwise, by the time the 
round has completed, the execution is already delayed passed the time for the 
maximum time elapse for the next 0-timed rewrite rule; i.e., the maximum speed 
of execution of the formal model and communication is slower than the real time 
requirements. Now, assuming that the constraint $t_{round} \leq mte(C_{next})$ 
is satisfied, there is still another problem we must deal with. The actual 
commands sent to the device are not received until $t_{jitter} = t_{comm1} + 
t_{rew1} + t_{comm2}$ time after they are actually supposed to be executed. This 
could be very problematic. Even if the model can keep up with real-time, the 
time in which it issues commands will be delayed. For example, the shocks from 
a pacemaker may be issued at the correct time by the model, but the real shock 
is not delivered until 0.1 seconds later. To meet this requirement, we need to 
look at the finer requirements of medical devices and patient parameters. How 
much jitter in control can a patient tolerate? As we have seen in the Section 
\ref{sec:case}, the jitter seems to be tolerable for the applications we 
considered, and furthermore, the end-to-end round communication timing 
constraints are also satisfied by our case studies.

%% file: conclusion.tex
\section{Conclusion}
\label{sec:conc}

Safety of medical devices and of their interoperation is an unresolved issue 
causing severe and sometimes deadly accidents for patients.  Formal methods, 
particularly in support of highly generic and reusable formal patterns whose 
safety properties have been verified can help in ensuring the safety of 
specific components, but this still leaves several open problems including: (i) 
how to pass from specifications to code and from logical time to physical time 
in a correctness-preserving ways; and (ii) how to experimentally validate 
medical safety architectures in realistic scenarios in which actual devices and 
models of patients and doctors can interact with formally specified and 
provably safe designs of device components.

By developing virtual emulation environments in which highly generic and 
reusable formally verified patterns in Real-Time Maude can be easily 
transformed into emulations in physical time which can interact with other real 
devices and with simulations of patient and/or doctor behaviors, we have taken 
some first steps towards a seamless integration of formal specification and 
verification with emulation and testing, and ultimately with deployment of 
medical DES systems that offer much stronger safety guarantees than what is 
currently available.  Much work remains ahead.  As we explain in 
\cite{sun-meseguer-sha-wrla10}, the provably safe formal pattern used in the 
experiments of this paper is just one such pattern: it covers a useful class, 
but does not cover other kinds of safety needed in other medical devices. Also, 
other safety concerns, such as so-called open-loop safety, ensuring that 
medical devices will always be in states safe for the patient even under key 
infrastructure failures, such as network disconnection, have not been addressed
in this work.  However, we believe that the general methodology presented here 
to pass from formal specifications to virtual emulation environments and 
eventually to deployed systems should also be applicable to those new formally 
verified patterns that have yet to be developed.

%% file: main.bbl
\begin{thebibliography}{10}

\bibitem{alur04}
Rajeev Alur and {et. al.}
\newblock {Formal specifications and analysis of the computer-assisted
  resuscitation algorithm (CARA) Infusion Pump Control System}.
\newblock {\em International Journal on Software Tools for Technology
  Transfer}, 2004.

\bibitem{arney07}
David Arney and {et. al.}
\newblock {Formal Methods Based Development of a PCA Infusion Pump Reference
  Model: Generic Infusion Pump (GIP) Project}.
\newblock In {\em Joint Workshop on HCMDSS/MDPnP}, 2007.

\bibitem{fullmaude}
F.~Dur{\'a}n and J.~Meseguer.
\newblock The {Maude} specification of {Full Maude}.
\newblock Technical report, SRI International, 1999.

\bibitem{HealthGrades2009}
HealthGrades.
\newblock {The Sixth Annual HealthGrades Patient Safety in American Hospitals
  Study}, 2009.

\bibitem{jetley06}
Raoul Jetley, S.~Purushothaman Iyer, and Paul~L. Jones.
\newblock A formal methods approach to medical device review.
\newblock {\em Computer}, 39:61--67, 2006.

\bibitem{rt-maude}
P.~C. {\"O}lveczky and J.~Meseguer.
\newblock Semantics and pragmatics of {Real-Time Maude}.
\newblock {\em Higher-Order and Symbolic Computation}, 20(1-2):161--196, 2007.

\bibitem{rtmaude-manual}
Peter {\"O}lveczky.
\newblock {Real-Time Maude 2.3 Manual}, August 2007.

\bibitem{ray04}
Arnab Ray and Rance Cleaveland.
\newblock {Unit verification: the CARA experience}.
\newblock {\em International Journal on Software Tools for Technology
  Transfer}, 2004.

\bibitem{tech-rep}
Mu~Sun, Jos\'{e} Meseguer, and Lui Sha.
\newblock {A Formal Pattern Architecture for Safe Medical Systems}.
\newblock https://netfiles.uiuc.edu/musun/www/medical\_pattern/techrep.pdf.

\bibitem{sun-meseguer-sha-wrla10}
Mu~Sun, Jos\'{e} Meseguer, and Lui Sha.
\newblock {A Formal Pattern Architecture for Safe Medical Systems}.
\newblock In {\em 8th International Workshop on Rewriting Logic and Its
  Applications (WRLA'2010)}, 2010.

\end{thebibliography}
